\documentclass[journal]{IEEEtran}
\usepackage{color,soul}
\usepackage{cite}
\usepackage{graphicx}
\usepackage{amsmath}
\usepackage[utf8]{inputenc}
\usepackage{amssymb}
\usepackage{pifont}
\usepackage{array}
\usepackage{lipsum}
\usepackage{multirow}
\usepackage[english]{babel}
\usepackage{float}
\usepackage[caption=false,font=footnotesize]{subfig}
\usepackage{algorithm} 
\usepackage{algorithmic}
\usepackage[table,xcdraw]{xcolor}
\usepackage{soul}
\usepackage[T1]{fontenc}
\usepackage[font=small,labelfont=bf,tableposition=top]{caption}

\DeclareCaptionLabelFormat{andtable}{#1~#2  \&  \tablename~\thetable}
\ifCLASSINFOpdf
 
\else
\fi

\begin{document}

\title{MERAM: Non-Volatile Cache Memory Based on Magneto-Electric FETs}
\author{\IEEEauthorblockN{
Shaahin Angizi\IEEEauthorrefmark{2},
Navid Khoshavi\IEEEauthorrefmark{1},
Andrew Marshall\IEEEauthorrefmark{3},
Peter Dowben\IEEEauthorrefmark{4} and
Deliang Fan\IEEEauthorrefmark{2} 
}

\IEEEauthorblockA{\IEEEauthorrefmark{2}School of Electrical, Computer and Energy Engineering, Arizona State University, Tempe, AZ 85287\\
\IEEEauthorrefmark{1}Department of Computer Science, Florida Polytechnic University, FL\\
\IEEEauthorrefmark{3}Department of Electrical and Computer Engineering, The University of Texas at Dallas, Richardson, TX\\
\IEEEauthorrefmark{4}Department of Physics and Astronomy,  University of Nebraska- Lincoln, Lincoln, NE\\
sangizi@asu.edu, nkhoshavinajafabadi@floridapoly.edu, Andrew.Marshall@utdallas.edu, pdowben@unl.edu, dfan@asu.edu} \vspace{-2em}}

\markboth{}%
{Shell \MakeLowercase{\textit{et al.}}: Bare Demo of IEEEtran.cls for IEEE Journals}

\maketitle
\vspace{-2em}
\begin{abstract}

Magneto-Electric FET (MEFET) is a recently developed post-CMOS FET, which offers intriguing characteristics for high speed and low-power design in both logic and memory applications.    
In this paper, for the first time, we propose a non-volatile 2T-1MEFET memory bit-cell with separate read and write paths. We show that with proper co-design at the device, cell and array levels, such a design is a promising candidate for fast non-volatile cache memory, termed as MERAM. To further evaluate its performance in memory system, we, for the first time, build a device-to-architecture cross-layer evaluation framework based on an experimentally-calibrated MEFET device model to quantitatively analyze and benchmark the proposed MERAM design with other memory technologies, including both volatile memory (i.e. SRAM, eDRAM) and other popular non-volatile emerging memory (i.e. ReRAM, STT-MRAM, and SOT-MRAM).
The experiment results show that MERAM has a high state distinguishability with almost 36x magnitude difference in sense current. Results for the PARSEC benchmark suite indicate that as an L2 cache alternative, MERAM reduces Energy Area Latency (EAT) product on average by $\sim$98\% and $\sim$70\%  compared with typical 6T SRAM and 2T SOT-MRAM platforms, respectively.

\end{abstract}


\IEEEpeerreviewmaketitle

\section{Introduction}
Over the past decade, Non-Volatile Memories (NVM) have been actively explored with the main goals of satisfying robustness, minimizing stand-by leakage, achieving high speed and integration density as major requirements to replace conventional volatile memory technologies in main memory (i.e. DRAM) or cache (i.e. SRAM) \cite{dowben2018towards,das2011fetram,dowben2020magneto}. This could optimistically boost the memory capacity and performance especially when it comes to on-chip cache for embedded applications and low-energy budget IoT systems. However, there is very few memory technologies still surviving in this arena. Among popular NVM technologies, ReRAM \cite{akinaga2010resistive} and PCM \cite{lee2009architecting} offer higher ON/OFF ratio, thus higher sense margin, and packing density than DRAM ($\sim$2-4$\times$) \cite{lee2009architecting}. However, they suffer from slow and power hungry write operations as well as low endurance ($\sim10^{5}$-$10^{10}$) \cite{dong2008circuit,dowben2020magneto}.
Recent experiments and fabrication of spin-based NVMs show the ability to switch the magnetization using current-induced Spin-Transfer Torque (STT) or Spin-Orbit Torque (SOT) with high speed (sub-nanosecond), long retention time (10 years) and less than $fJ/bit$ memory write energy (close to SRAM) \cite{angizi2018cmp,angizi2018imce}. However, such NVMs have poor ON/OFF ratios (maximum resistance ratio $\sim$7$\times$) in parallel and anti-parallel configurations. Moreover, the current densities in current-driven spin devices impose reliability issues and large static power dissipation \cite{angizi2019mrima,angizi2018pima}.
The ferroelectric transistor random access memories (FERAMs) \cite{das2011fetram} offer high endurance and sense margin, but suffering from a destructive read operation. FE-FET memories \cite{george2016nonvolatile,reis2018computing}, however, offer FERAMs' benefits with a reduced 1-10 ns \cite{george2016nonvolatile} write time and could be a possible alternative. The downside is their large write voltage ($>4.0 V$) and power consumption \cite{ni2018circuit,reis2019design,DayaneGLS}

Recently, another promising spintronic device, based on Magneto-Electric (ME) phenomena \cite{dowben2018towards,sharma2017verilog,sharma2020evolving,dowben2020magneto} has shown superior performance in terms of switching speed, energy, ON/OFF ratio, etc.
The principle innovative feature of this emerging device that significantly differs from the traditional spintronic devices is that
% control of the magnetic state of a free ferromagnetic layer by means of exchange bias produced by the ME’s boundary magnetization. Such a scheme averts the complexity and detrimental switching energies associated with ME exchange-coupled ferromagnetic devices, instead being based on the switching of a ME. As a result, 
switching speed is only limited  by the switching dynamics of the ME material of the voltage controlled spintronic devices \cite{dowben2018towards,nikonov2015benchmarking,parthasarathy2018reversal}. 
With coherent rotation, as the domain switching mechanism, the switching speed might be as fast as 5-6 ps \cite{parthasarathy2018reversal} as it doesn't require the switching of a ferromagnet or movement of a ferromagnetic domain wall. Therefore, this may be considered to be spintronics without a ferromagnet, achieving fast write speeds ($<$20 ps/full-adder) \cite{parthasarathy2018reversal,sharma2017novel,nikonov2015benchmarking}, a low energy cost ($<$20 aJ/full-adder) \cite{sharma2017novel}, combined with a great temperature stability (operational to 400 K or more), and improved scalability.

While there are proposals for logic design based on ME devices, such as ME-MTJ \cite{sharma2017novel,sharma2020circuits} and MEFET \cite{sharma2017verilog,dowben2018towards,nikonov2015benchmarking,pan2018complementary}, etc., to the best of our knowledge, there is no published ME memory design and evaluation. 
In this work, we are the first to propose a 3-terminal MEFET-based non-volatile memory cell design with separate read and write paths. We show that with proper co-design at the device, cell and array levels, such design may be a potential competitor for current non-volatile on-chip cache replacement race. 
Our main contributions in this work are summarized as follows:
\begin{itemize}
    \item We enhance the experimentally-calibrated MEFET Verilog-A circuit model \cite{sharma2018compact} to perform extensive analysis at the device, cell/array circuit levels.
    \item  We propose a 2-transistor 1-MEFET memory bit-cell circuit design with high speed and low read/write energy suitable for on-chip memory cache.
    \item We develop a bottom-up evaluation framework to extensively assess and benchmark MEFET cache performance with current volatile and non-volatile memories, including SRAM, eDRAM, ReRAM, STT-MRAM, SOT-MRAM.
\end{itemize}
 \vspace{-1.1em}

\section{Magneto-Electric Spin Field Effect Transistor (MEFET)}
\subsection{Device Characterization}

\begin{figure}[h]
\begin{center}
\begin{tabular}{c}
\includegraphics [width=0.99\linewidth]{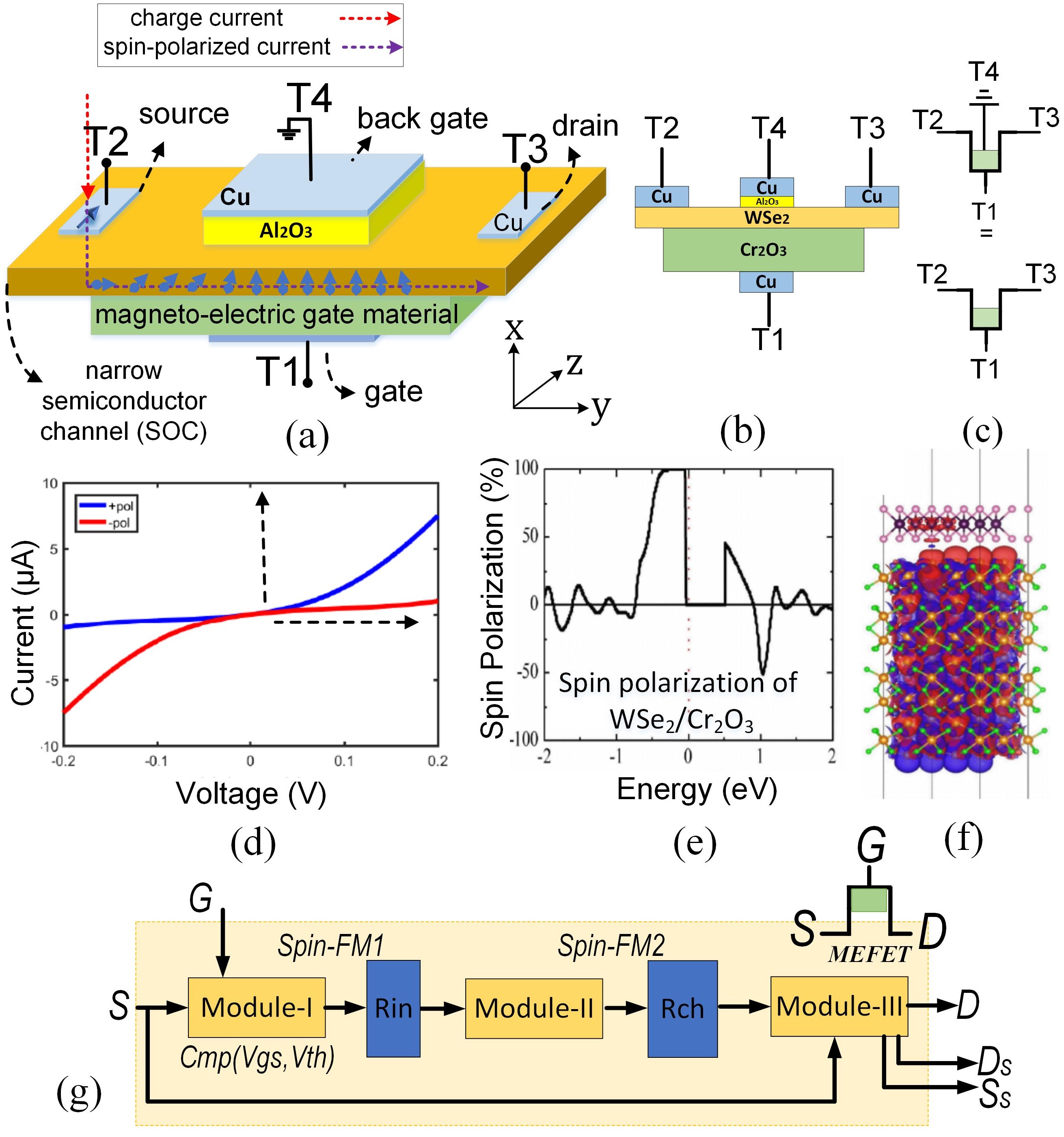}\vspace{-0.4em}
 \end{tabular} \vspace{-0.7em}
\caption{(a) The basic Magneto-Electric spin-FET (MEFET) with gate, source, drain and back gate. The narrow  semiconductor channel can be made of any suitable material (e.g: graphene, InP, GaSb, PbS, WSe\textsubscript{2}, etc.), (b) A 2D view of MEFET, (c) The proposed MEFET circuit scheme, (d) Source to drain current versus voltage at T1 in the ME-FET. The SOC channel polarized in opposite directions (+ or -) by the ME gate, (e) Induced spin polarization in WSe\textsubscript{2}, and (f) Interaction with chromia adapted from \cite{dowben2018towards}, (g) Verilog-A modules developed for MEFET modeling.}\vspace{-1em}
\label{MEFET_device}
\end{center}
\end{figure}

The Magneto-Electric spin Field Effect Transistor (MEFET) is structurally very similar to the conventional CMOS FET device. 
Fig. \ref{MEFET_device}a shows the basic single source version of MEFET as a 4-terminal device with gate (at T1), source (T2), drain (T3), and back gate (T4) terminals \cite{chuang2016low}. The device is a stacked structure of a narrow semiconductor channel sandwiched by two dielectrics i.e. Magneto-Electric (ME) material (e.g. Chromia (Cr\textsubscript{2}O\textsubscript{3})) and insulator (e.g. Alumina (Al\textsubscript{2}O\textsubscript{3})). 
There are two electrodes contacting the stacked structure, at the bottom gate via ME layer (T1) and at the top via back gate alumina layer. The channel as shown in Fig. \ref{MEFET_device}b is made of tungsten diselenide (WSe\textsubscript{2}), enabling on-off ratio of $\sim10^6$ \cite{chuang2016low} and high hole mobility comparable to CMOS. The source (at T2 terminal) could be made of both fixed spin ferro-magnetic (FM) polarizer or a conductor. MEFET operates based on programming of the semiconductor channel polarization so-called Spin Orbit Coupling (SOC), by the boundary polarization of the ME gate, through the proximity effect. In other words, the channel can be polarized by the ME layer on extremely low voltage of around $\pm$100 mV \cite{dowben2018towards,sharma2017verilog,sharma2020evolving} at T1 while T4 is grounded. 

The ME layer has high interface polarization which can be controlled by vertical voltage \cite{dowben2018towards}. Chromia here is a promising ME gate dielectric that has the potential to induce spin polarization in an over-layer channel \cite{dowben2018towards}. Applying voltage across the gate and back gate terminal is equivalent to charging of the ME capacitor. Therefore, depending on the positive or negative voltage applied to T1, a vertical electrical field across the gate is created. In response to the electrical field, paraelectric polarization and Atomic
Force Microscopy (AFM) order in ME insulator layer are switched. It first changes the direction of orientation of chromia spin vectors through SOC.
The ME boundary polarization can have an exchange interaction with a semiconductor channel to polarize the carriers' spins in the channel and induce preferred conduction, i.e., much lower resistance, in only one direction along the channel. 
This high spin boundary polarization was predicted independently by Andreev and Belashchenko \cite{andreev1996macroscopic,belashchenko2010equilibrium} and has been experimentally confirmed, for ME chromia, by a wide variety of techniques \cite{dowben2018towards}. 
In other words, the influence of surface magnetization on the channel produces directionality of conduction, which is not possible through conventional gate dielectrics, as depicted in Fig. \ref{MEFET_device}d. The current versus voltage dependent on the direction of ME polarization is obtained by NEGF transport simulation \cite{anantram2008modeling} in a 2-D ribbon with a width of 20 nm and band mass of 0.1$m_e$. We consider a conservative value of exchange splitting of 0.1 eV, T3-T2 = 0.1 V, at 300 K. As show in Fig. \ref{MEFET_device}e,  chromia induces a very high level of spin polarization in WSe\textsubscript{2} channel, virtually 100\% at the top of the valence band for hole conduction. Such interaction with chromia is shown in Fig. \ref{MEFET_device}f (adapted from author's preliminary work in \cite{dowben2018towards}).
Therefore, in the end, the channel spin vector changes to either ‘up’ or ‘down’ direction, as shown in Fig. \ref{MEFET_device}d.  
After biasing the SOC channel, the charge current is injected through source generating a spin-polarized current at the T3.  Fig. \ref{MEFET_device}b and Fig. \ref{MEFET_device}c show the 2D view of the MEFET device and its transistor circuit representation used in this work. As T4 is grounded, the simplified three-terminal scheme is used hereafter.

Compared to a 200 ps coupling delay of the Magneto-Electric Magnetic Tunnel Junction (ME-MTJ) \cite{sharma2017novel}, MEFET achieves an extremely low switching delay (somewhere in the region of 10-100 ps \cite{manipatruni2015spin}, thus avoiding the excessive delays associated with exchange-coupled ferromagnets. Such device shows the feature of non-volatility due to the non-volatile AFM ordering of the ME and a very high and sharp turn-on voltage due to the sharp turn on of the ME-switching \cite{sharma2018compact}. 
It it worth pointing out that for sensing of device resistance, a more energy-efficient read circuitry could be expected for  MEFET as it shows much higher ON/OFF ratio compared with TMR-sensing in traditional spin-based devices. The ON/OFF current ratio for WSe\textsubscript{2} \cite{fang2012high} is experimentally shown to extend up to $10^6$, while magneto-resistance effect in MTJs does not exceed $10^2$. 
Regarding MEFET fabrication status, there exists many experiments proving that with a static magnetic field, the chromia's magnetic order could be switched back and forth via applying an electric field \cite{kosub2017purely,toyoki2015magnetoelectric}. However, the creation of a desired direction via SOC is yet to be proved experimentally and has been only anticipated based on principle computations  \cite{dowben2018towards,pan2018complementary}. 

\vspace{-0.7em}

% \begin{figure}[t]
% \begin{center}
% \begin{tabular}{c}
% \includegraphics [width=0.88\linewidth]{./Figures/model.pdf}\vspace{-0.4em}
%  \end{tabular} \vspace{-0.2em}
% \caption{(a) Single source MEFET modeling scheme. R\textsubscript{in} and R\textsubscript{ch} represent the internal and channel resistance
% of the device, respectively. (b) }\vspace{-0.7em}
% \label{model}
% \end{center}
% \end{figure}

\subsection{Device Modeling}
We consider two aspects in developing the MEFET model: First, ME control of the channel spin polarization based on the proximity induced polarization in the narrow 2D channel. Second, spin injection/detection function at the source/drain \cite{sharma2018compact,sharma2017verilog}. The model is developed in Verilog-A with three distinct modules as shown in Fig. \ref{MEFET_device}g. In the presented MEFET model, T2 and T3 are utilized to inject the spin-polarized current and detect it, respectively.

% The correlation between detected voltage and injected current, so-called spin-efficiency, can be achieved similar to the spin-FET \cite{wang2017modeling} as:

% \begin{equation}
%   \small  \frac{V_{detect}}{I_{inject}}=\pm \frac{1}{2}P_J^{2}R_Ne^{-\frac{L}{\lambda_N}}
% \end{equation}

% Where, $\pm$ sign shows the directions of the opposite voltages, depending on the direction of the FM magnetization. $P_J$ represents the interfacial current polarization showing the polarization capacity of the junctions. $R_N$ is the resistance of SOC channel. $L$ is the channel length and $\lambda_N$ denotes the spin-diffusion length of SOC layer, which depends on the property of the channel material. Now, considering the Datta–Das theory, the MEFET can be modeled as:

% \begin{equation}
% \small V_{out}=\pm V_{detect}\times cos\left ( \frac{2m^*\times \alpha (V_G)\times L}{\hbar^2 }+\varphi \right )
% \end{equation}

% Here, $m^*$ is the electron's effective mass, $\alpha$ is the SOC coefficient as a function of gate voltage, $\hbar$ denotes the reduced Planck’s constant, and $\varphi$ is the phase shift correction parameter. 

In the Verilog-A modeling shown in Fig. \ref{MEFET_device}g, we further consider the experimental switching parameters for chromia layer and SOC channel. Module-I receives the gate and source voltage and compares the gate-source voltage with the threshold of chromia state inversion (0.050V \cite{sharma2017verilog}), initializes the memory and assigns back to voltage across the drain and source terminal. Module-II considers the delay factor in transition between source to drain. We consider a computed delay element is associated with the boundary magnetization between the ME film and the interface of channel. The switching time of MEFET device is then limited only by the switching dynamics of ME. Module-III assigns the proper R\textsubscript{ch} and calculates corresponding electrical parameters output voltage at the drain. The channel resistance (R\textsubscript{ch}) across the two-dimensional (2D) narrow channel is considered in series to the input resistance R\textsubscript{in} to define the boundary conditions for switching. Besides, we consider two spin state terminals (“Ss” and “Ds”)  to validate the spin state injected/detected at the source/drain terminals as shown in Fig. \ref{MEFET_device}g. The ‘up’ and ‘down’ spins are represented by constant voltages sources with ‘+1 V’ and ‘-1 V’, respectively at the “Ss” terminal. Before running the simulation, the injected spin orientation can be selected, making the model flexible to be used in various CAD tools such as Cadence platform.
% The highly resistive \cite{kwan2015electric} ME layer induces spin polarization in the channel due to the proximity of magnetic atoms or a magnetically ordered substrate \cite{dowben2018towards}.
The spin current then can be detected at the drain (“Ds” in Fig. \ref{MEFET_device}g). In our compact model, the processional delay across the FM layer was taken into account by a fixed delay assumed to be 200 ps. This assumption is based on the best estimate of the coupling delay \cite{nikonov2015benchmarking}. Table \ref{param} lists the parameter values used in the models. \vspace{-0.5em}

\begin{table}[t]
\centering
\caption{Compact model parameters of the MEFET, used in the Verilog-A model adapted from \cite{sharma2017novel,sharma2018compact}}\vspace{-0.8em}
\scalebox{1}{
\begin{tabular}{|c|c|c|}
\hline
Parameter        & Value & Description of Parameter and Units                            \\ \hline
$\epsilon_{ME}$  & 12    & Dielectric constant of chromia \cite{iyama2013magnetoelectric}                        \\ \hline
$\epsilon_{Al_2O_3}$ & 10    & Dielectric constant of Alumina       \\ \hline
$t_{ME}$         & 10    & thickness of magnetoelectric layer, nm     \\ \hline
$W_{ME}\times L_{ME}$   & 900  & area of magnetoelectric layer, $nm^2$         \\ \hline
$t_{ox}$         & 2     & Oxide barrier thickness, nm         \\ \hline
$V_t$            & 0.05   & Threshold of Chromia state inversion, V    \\ \hline
$V_g$            & 0.1   & Voltage applied across ME layer, V    \\ \hline
$R_{on}$            & 1.05   & ON Resistance, $k\Omega$     \\ \hline
$R_{off}$            & 63.4   & OFF Resistance, $M\Omega$     \\ \hline
% TMR              &    & Tunnel magnetoresistance     \\
\end{tabular}} \vspace{-1em}
\label{param}
\end{table}

%  \begin{figure}[t]
% \begin{center}
% \begin{tabular}{c}
% \includegraphics [width=0.88\linewidth]{./Figures/bitcell2.pdf}\vspace{-0.4em}
%  \end{tabular} \vspace{-0.7em}
% \caption{(a) The 2T-1MEFET RAM bit-cell (a) Read signals, (b) Write signals.}\vspace{-1.8em}
% \label{MEFET}
% \end{center}
% \end{figure}

\section{MEFET Memory}
\begin{figure}[t]
\begin{center}
\begin{tabular}{c}
\includegraphics [width=1.01\linewidth]{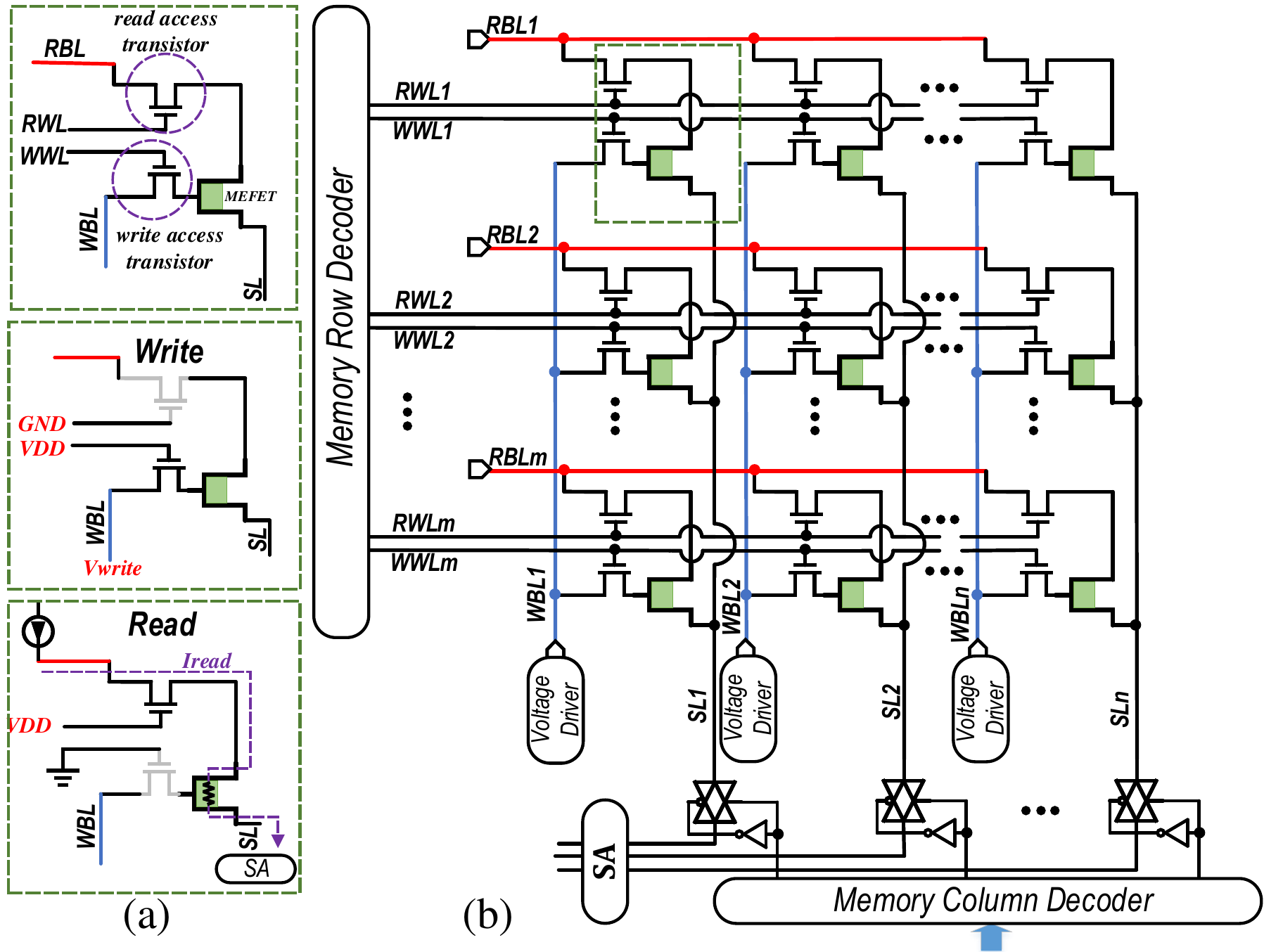}\vspace{-0.4em}
 \end{tabular} \vspace{-0.7em}
\caption{(a) The 2T-1MEFET RAM (MERAM) bit-cell with Read and Write signals, (b) An m$\times$n MERAM array with peripheral circuitry.}\vspace{-1.3em}
\label{array}
\end{center}
\end{figure}

The proposed non-volatile 2T-1MEFET RAM bit-cell, called MERAM, consists of one MEFET as the main storage element and two access transistors, as shown in Fig. \ref{array}a. By virtue of three-terminal structure of MEFETs, depicted in Fig. \ref{MEFET_device}c, we design the memory bit-cell to have separate read, write paths which facilitates independent optimization of both operations, and avoids read-write conflicts in many other 1T1R resistive non-volatile memory designs. Each cell is controlled by five controlling signals i.e. Write Word Line (WWL), Write Bit Line (WBL), Read Word Line (RWL), Read Bit Line (RBL), and Source Line (SL). The read/write access transistor is controlled by RWL/WWL enabling selective read/write operation on the cells located within one row. An m$\times$n MERAM array developed based on the proposed bit-cell is shown in Fig. \ref{array}b.
The WLs and RBLs are shared amongst the cells within the same row and WBLs and SLs are shared between cells in the same column. The WLs are controlled by the memory row decoder (active-high output). The column decoder (active-high output) controls the activation of read current path through the SL. The voltage driver component is designed to set the proper voltage on the WBL. In the following, we explain the read and write operations, respectively.

\subsection{Read Operation}
\begin{figure}[t]
\begin{center}
\begin{tabular}{c}
\includegraphics [width=0.99\linewidth]{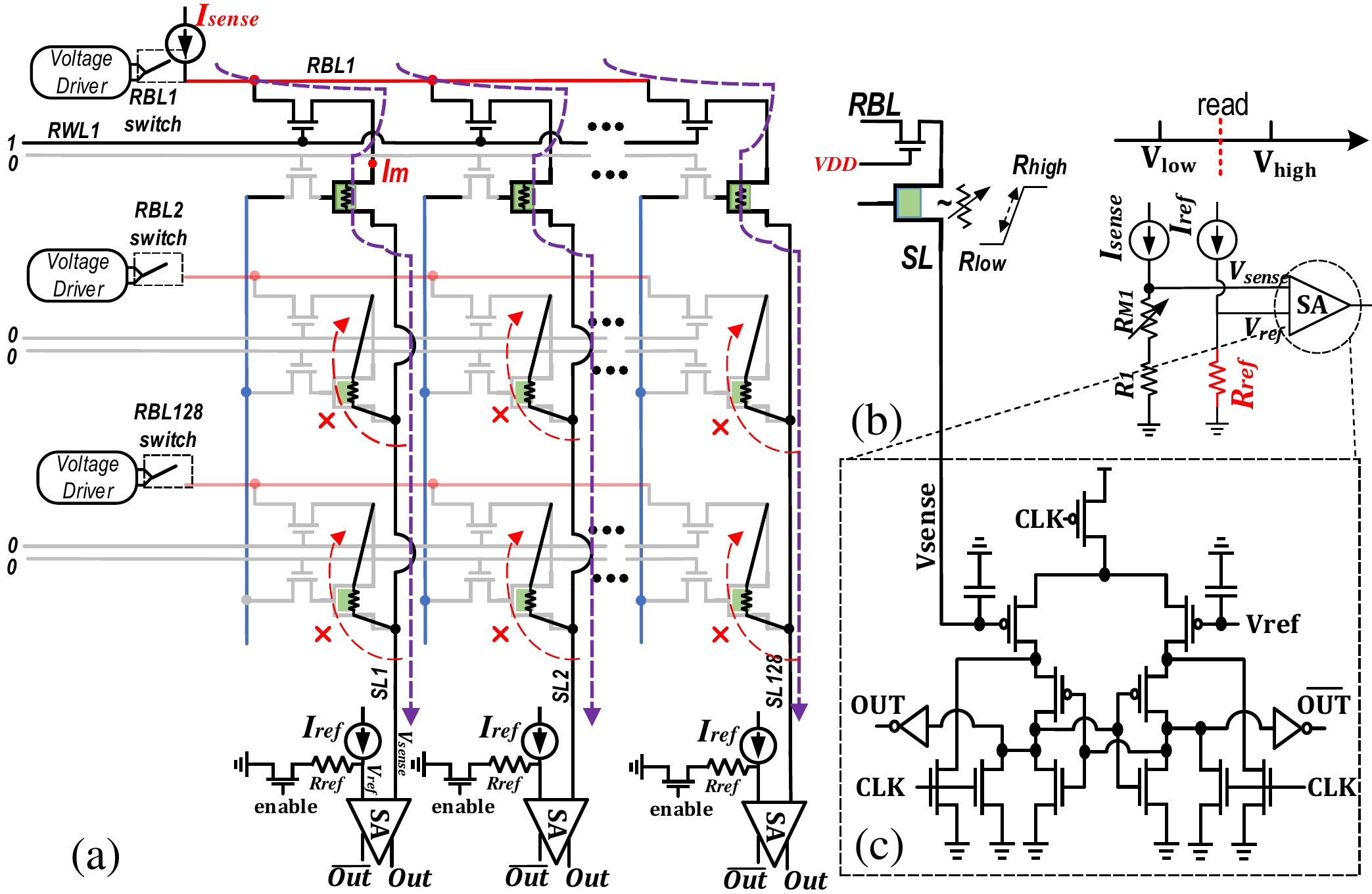}\vspace{-0.4em}
 \end{tabular} \vspace{-0.2em}
\caption{(a) The array-level read operation. It can be seen that read access transistors isolate the un-accessed cells avoiding sneak paths. (b) The idea of voltage comparison between V\textsubscript{sense} and V\textsubscript{ref} for memory read, (c) The Sense Amplifier (SA).} \vspace{-1.3em}
\label{array2}
\end{center}
\end{figure}

The concept behind MERAM's read operation is to sense the resistance of the selected memory cell and compare it by a reference resistor using a Sense Amplifier (SA), as shown in Fig. \ref{array}a. At the array level, shown in Fig. \ref{array2}a, the row and column decoders activate the RWL and SL paths respectively. When a memory cell is selected, by applying a very small sense current (sub-micro) to RBL, a voltage (V\textsubscript{sense}) is generated on the corresponding SL, which is taken as the input of the sense circuit, as shown in Fig \ref{array2}b. Owing to the low or high resistance state of the selected 2T-1MEFET RAM bit-cell (R\textsubscript{M1}), the sense voltage is V\textsubscript{low} or V\textsubscript{high} (V\textsubscript{low}$<$V\textsubscript{high}), respectively. Thus, through setting the reference voltage at (V\textsubscript{AP}+V\textsubscript{P})/2, the SA outputs binary `1' when V\textsubscript{sense}$>$V\textsubscript{ref}, whereas output is `0'. We designed and tuned the sense circuit based on StrongARM latch \cite{razavi2015strongarm} shown in Fig. \ref{array2}c. Each read operation requires two clock phases: pre-charge (CLK `high') and sensing (CLK `low'). During the pre-charge phase, both SA's outputs are reset to ground potential. Then, in the sensing phase, the input transistors provide various charging current based on the gate biasing voltages (V\textsubscript{sense} and V\textsubscript{ref}), leading to various switching speed for the latch's cross-coupled inverters. The biasing condition for the read operation is tabulated in Table \ref{condition}.

\subsection{Write Operation}
The write operation, shown in Fig. \ref{array}a, is accomplished by activating  WWL and asserting appropriate bipolar write voltage ($V_{write}$/-$V_{write}$) through voltage driver on WBL. As detailed above, the voltage provides enough vertical electrical field magnitude across the gate to switch the spin vectors of underlying chromia layer and channel of MEFET. The RBL and SL don't require to be grounded for the  write period as seen in other memory technologies such as FEFET \cite{reis2019design}. The unaccessed rows are isolated by driving the corresponding WLs to GND by row decoder. This is necessary to avoid unwanted current paths that can cause false write/read-states. The biasing condition for the write operation is tabulated in Table \ref{condition}.\vspace{-1em}

\begin{table}[h]
\centering
\caption{Bias configuration of MERAM array.}\vspace{-0.5em}
\scalebox{0.87}{
\begin{tabular}{|c|c|c|c|c|c|c|}
\hline
           & Operation                                                                   & WWL & RWL & WBL                      & RBL        & SL         \\ \hline
Accessed   & \multirow{2}{*}{Read}                                                       & 0   & 1   & -                        & $I_{sense}$ & - \\ \cline{1-1} \cline{3-7} 
Unaccessed &                                                                             & 0   & 0   & -                        & -          & -          \\ \hline
Accessed   & \multirow{2}{*}{\begin{tabular}[c]{@{}c@{}}Write \\ ('1'/'0')\end{tabular}} & 1   & 0   & - $V_{write}$/-$V_{write}$ & -          & -          \\ \cline{1-1} \cline{3-7} 
Unaccessed &                                                                             & 0   & 0   & -                        & -          & -          \\ \hline
All        & Hold                                                                        & 0   & 0   & -                        & -          & -          \\ \hline
\end{tabular}}
\label{condition}
\end{table}

% \begin{figure}[t]
% \begin{center}
% \begin{tabular}{c}
% \includegraphics [width=0.69\linewidth, height=6cm]{./Figures/MEFET_write.pdf}\vspace{-0.4em}
%  \end{tabular} \vspace{-0.7em}
% \caption{The array-level write operation.} \vspace{-2.3em}
% \label{array3}
% \end{center}
% \end{figure}

\section{Evaluation}

\subsection{Bottom-Up Evaluation Framework}

\begin{figure}[t]
\begin{center}
\begin{tabular}{c}
\includegraphics [width=0.99\linewidth]{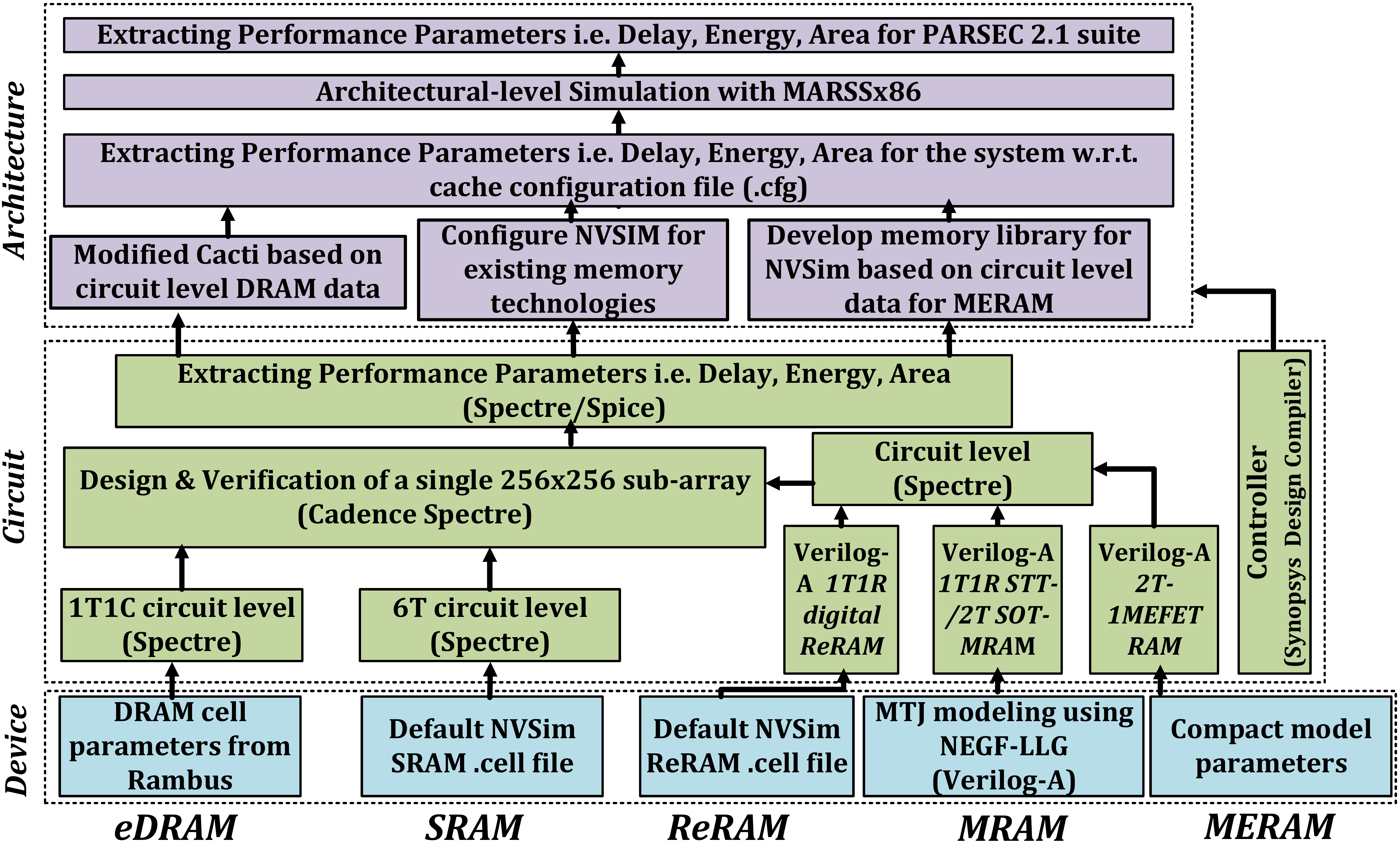}\vspace{-0.4em}
 \end{tabular} \vspace{-0.2em}
\caption{The bottom-up evaluation framework developed for cache memory evaluation.}\vspace{-1.8em}
\label{flowchart}
\end{center}
\end{figure}

For cross-technology evaluation and comparison, we developed a comprehensive bottom-up cross-layer framework as shown in Fig. \ref{flowchart}. 
The device-level model of each technology was first developed/extracted from various models and assessments. For MERAM, we used the Verilog-A model developed by our co-author Dr. Marshall in preliminary work \cite{sharma2017verilog}. For STT-MRAM and SOT-MRAM, we jointly use the Non-Equilibrium Green's Function (NEGF) and Landau-Lifshitz-Gilbert (LLG) equations to model the bitcell, developed in our co-author Dr. Fan's preliminary work \cite{fong2016spin,angizi2019accelerating,angizi2019graphs}. We leveraged the default ReRAM and SRAM cell configuration of NVSim \cite{dong2012nvsim}. The eDRAM cell parameters were adopted and scaled from Rambus \cite{Rambus}. 
In the circuit-level, we developed a 256$\times$256 memory sub-array for each memory technology with peripheral circuity, simulated in Cadence Spectre with the 45nm NCSU Product Development Kit (PDK) library \cite{NCSU_PDK}.
At the architecture-level, we developed memory libraries for MERAM in C++, based on circuit level results and configure other memory technologies based on existing NVSim's \cite{dong2012nvsim} and Cacti's \cite{thoziyoor2008cacti} libraries. Performance data (i.e. delay, energy, area) for cache are estimated with respect to a single input memory configuration file (.cfg) as tabulated in Table \ref{ISOCAP}. The results are then fed to the cycle-accurate  MARSSx86 simulator \cite{patel2011marss} for each memory technology to show the architecture-level performance.
\vspace{-0.8em}

\begin{table*}[ht]
\begin{minipage}[b]{0.56\linewidth}
\centering
\scalebox{0.7}{
\begin{tabular}{|c|c|c|c|c|c|c|}
\hline
Metrics                         & ReRAM                  & STT-MRAM                & SOT-MRAM                & SRAM      & eDRAM         & MERAM         \\ \hline
Non-volatility                  & Yes                    & Yes                     & Yes                     & No        & No            & Yes           \\ \hline
\# of access transistors        & 1                      & 1                       & 2                       & 6         & 1             & 2             \\ \hline
Area ($mm^2$)                     & 1.77                   & 5.42                    & 5.85                    & 12.4      & 4.46          & 6.94          \\ \hline
Cache Hit Latency (ns)          & 2.55                   & 3.14                    & 5.07                    & 1.59      & 3.1           & 0.65          \\ \hline
Cache Miss Latency (ns)         & 1.21                   & 1.28                    & 1.32                    & 0.34      & -              & 0.22          \\ \hline
Cache Write Latency (ns)        & 20.5                   & 10.7                    & 3.93                    & 0.78      & 3.1           & 0.94          \\ \hline
Cache Hit Dynamic Energy (nJ)   & 0.33                   & 0.52                    & 0.21                    & 0.73      & 0.24          & 0.22          \\ \hline
Cache Miss Dynamic Energy (nJ)  & 0.033                  & 0.044                   & 0.03                    & 0.017     & -              & 0.037         \\ \hline
Cache Write Dynamic Energy (nJ) & 0.82                   & 1.27                    & 0.27                    & 0.72      & 0.24          & 0.27          \\ \hline
Cache Total Leakage Power (W)   & 0.38                   & 0.79                    & 0.21                    & 6.2       & 0.57          & 0.19          \\ \hline
Endurance                       & $\sim10^{5} - 10^{10}$ & $\sim10^{10} - 10^{15}$ & $\sim10^{10} - 10^{15}$ & Unlimited & $\sim10^{15}$ & $\sim10^{17}$ \\ \hline
Data Over-written Issue         & No                     & No                      & No                      & No        & Yes           & No            \\ \hline
\end{tabular} }
    \caption{Estimated row Performance of various memory technologies as a 4MB unified L2 cache with 64 Bytes cache line size.}
    \label{ISOCAP}
\end{minipage}\hfill
\begin{minipage}[b]{0.4\linewidth}
\centering
\includegraphics [width=0.92\linewidth]{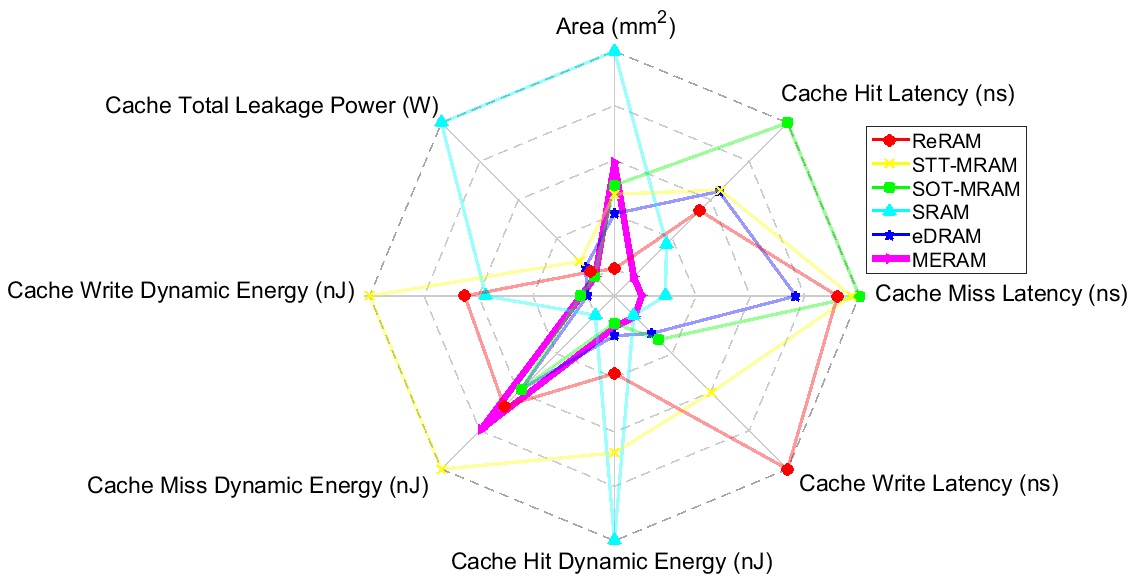}
\captionof{figure}{Benchmarking radar plot of MEFET vs. other technologies.}
\label{radar} \vspace{-2em}
\end{minipage}
\end{table*}

%%%%%%%%%%%%%%%%%%%%%%%%5555555

\subsection{Device and Circuit Level}

\begin{figure}[t]
\begin{center}
\begin{tabular}{cc}
\includegraphics [width=0.95\linewidth, height=8.5cm]{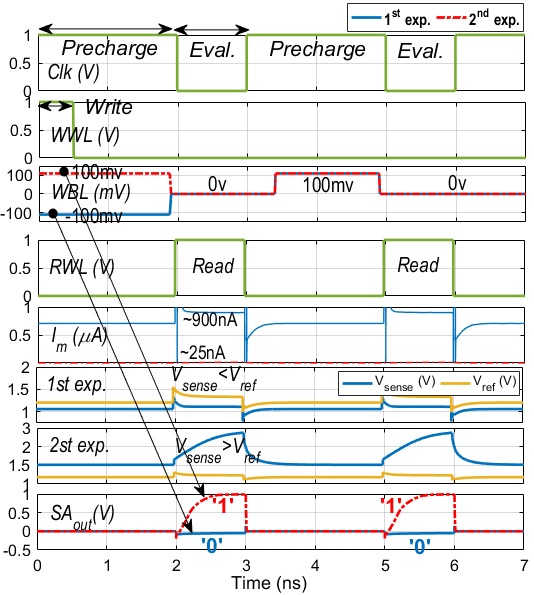}  \\ %[\abovecaptionskip]
 \end{tabular}\vspace{-0.8em}
\caption{The transient simulation results of two experiments on MERAM cell.}
\label{simulation}\vspace{-1.8em}
\end{center}
\end{figure}

Fig. \ref{simulation} shows the transient simulation results of a 2T-1MEFET RAM cell located in a 256$\times$256 sub-array based on the architecture shown in Fig. \ref{array}. Here, we consider two experiment scenarios for the write operation, as indicated by the solid blue and dotted red line in Fig. \ref{simulation}.
For the sake of clarity of waveforms, we assume a 3ns period clock synchronises the write and read operation. However,  $<$1ns period can be used for a reliable read operation.
During the precharge phase of SA (Clk=1), the V\textsubscript{write} voltage is set (=-100mv in the 1st experiment or +100mv in the 2nd experiment) and applied to the WBL to change the MEFET resistance to R\textsubscript{low}=1.05 $k\Omega$ or R\textsubscript{high}=63.4 $M\Omega$. Prior to the evaluation phase (Eval.) of SA, WWL and WBL is grounded while RBL is fed by the very small sense current, I\textsubscript{sense}= 900 $nA$. In the evaluation phase, RWL goes high and depending on the resistance state of MERAM bit-cell and accordingly SL, V\textsubscript{sense} is generated at the first input of SA, when V\textsubscript{ref} is generated at the second input of SA. The comparison between V\textsubscript{sense} and V\textsubscript{ref} for both experiments is plotted in Fig. \ref{simulation}.
We observe when V\textsubscript{sense}$<$V\textsubscript{ref} (1st experiment), the SA outputs binary `0', whereas output is `1' (2nd experiment). As can be seen, the same value can be sensed again in the next sensing cycle (3-6 ns) regardless of WBL voltage since WWL is deactivated so the MERAM bit-cell remains unchanged.

Table \ref{ISOCAP} compares the row performance of six various memory technologies, i.e. volatile SRAM and eDRAM with non-volatile ReRAM, STT-MRAM, SOT-MRAM, and MERAM integrated as a 4MB L2 cache with 64 Bytes cache line size. It is worth pointing out that the data is extracted from our bottom-up evaluation framework and architecture level simulators. Here, we discuss the results reported in Table \ref{ISOCAP}. Additionally, the radar plot in Fig. \ref{radar} further investigates and intuitively shows the pros and cons associated with each technology compared with MERAM in the array-level that could be employed to build high-density arrays.

\subsubsection{Latency}
As listed in Table \ref{ISOCAP}, we observe that MERAM shows a remarkable improvement in cache hit and miss latency (0.65/0.22 ns) as compared with other platforms. In terms of cache write, MERAM achieves the second shortest latency after volatile SRAM. It can achieve 0.94 ns cache write operation, which is $\sim$4$\times$ shorter than the best non-volatile memory (SOT-MRAM-3.93 ns) but not faster than SRAM counterpart (0.78 ns). Therefore, MERAM meets the requirements of a high speed non-volatile cache unit.

\subsubsection{Energy Consumption}
The energy budget for various cache operations is shown in Table \ref{ISOCAP}. The proposed MERAM design achieves a comparable dynamic energy consumption for cache hit to eDRAM and SOT-MRAM as the most energy-efficient designs. While SRAM, SOT-MRAM and ReRAM achieve the least energy consumption for cache miss. Moreover, we observe SOT-MRAM and MERAM platforms consume the smallest cache write dynamic energy among all the NVM platforms, due to their intrinsically low-power device operation, however eDRAM shows the least energy consumption. MERAM consumes 0.27 nJ for write operation, which is  $\sim$2.6$\times$ smaller than SRAM platform. MERAM writing technique averts dissipative currents and is thus energy-efficient and inerts against detrimental impacts of Joule heating \cite{dowben2018towards}.
The cache total leakage power is also reported in Table \ref{ISOCAP}. We observe that MERAM and SOT-MRAM consume the least leakage power compared to other candidates. 
Thus, MERAM could be considered as a promising non-volatile unit in terms of energy-efficiency.

\subsubsection{Endurance}
The inorganic MEFET easily lasts to $10^{17}$ switches \cite{sharma2017verilog}. The reason here is the required current densities are very low, which considerably reduces the device failure rate from what has been seen in other spintronic devices, which require much higher current densities ($10^{10}$-$10^{15}$) \cite{fukami2016spin}.

\subsubsection{Area}
The MERAM bit-cell requires two minimum size access transistors (W:L=90:50 at 45nm technology node) to enable separate read and write paths. In this way, MERAM occupies 6.94 mm$^2$ to implement a 4MB cache, which turns out to be a larger chip area  compared to other non-volatile memories and eDRAM. However, it  achieves  $\sim$1.7$\times$ smaller area than 6T SRAM platform. Therefore, MERAM array couldn't be considered an area-efficient non-volatile memory candidate compared to ReRAM, STT-MRAM and SOT-MRAM designs.

\subsubsection{Integration with CMOS} 
\begin{figure}[t]
\begin{center}
\begin{tabular}{c}
\includegraphics [width=0.89\linewidth]{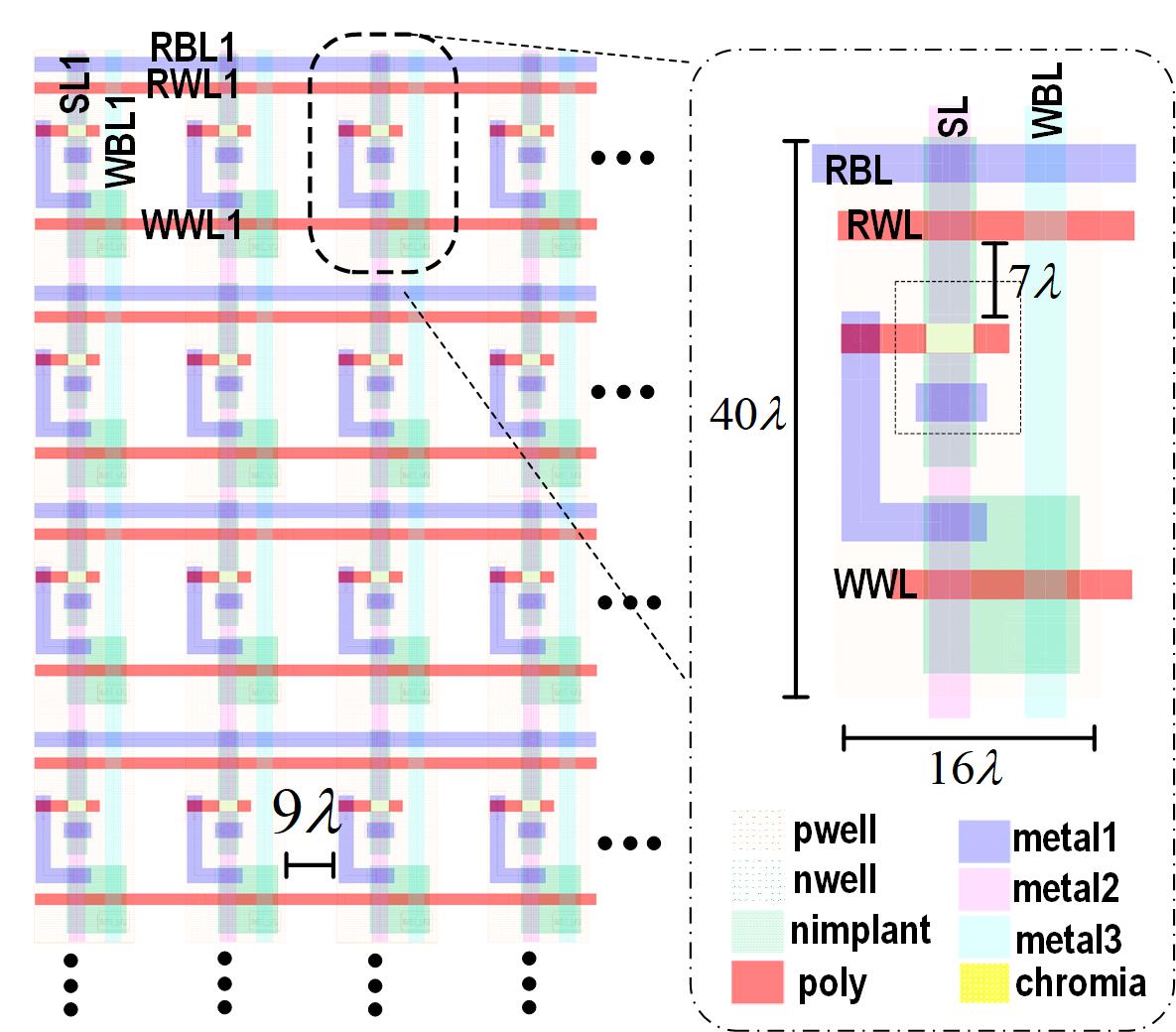}\vspace{-0.4em}
 \end{tabular} \vspace{-0.3em}
\caption{Layout of (a) 4$\times$4 MERAM array and (b) MERAM bit-cell.}\vspace{-1em}
\label{area}
\end{center}
\end{figure}

The device characteristics of chromia layer makes this material interesting for the integrating into the back end of line (BEoL). The feasibility of integration of chromia with silicon was experimentally demonstrated in \cite{punugupati2014strain,panda2018crystallographic}.
The bit-cell layout of single-port MERAM, shown in Fig. \ref{area}(b), was estimated using $\lambda$-based layout rules ($\lambda$: half of the minimum feature size, F, here $\lambda$:22.5nm) \cite{gupta2012layout}. The proposed cell takes estimately 40$\lambda \times$16$\lambda$=640$\lambda^2$, in contrast to 2000$\lambda^2$ of layout of baseline 6T-SRAM adapted from \cite{guthaus2016openram}. The layout of 4$\times$4 MERAM array with controlling signals is illustrated in Fig. \ref{area}(a).

% \begin{figure}[b]
% \begin{center}
% \begin{tabular}{c}
% \includegraphics [width=0.97\linewidth]{./Figures/Radar.jpg}\vspace{-0.4em}
%  \end{tabular} \vspace{-0.3em}
% \caption{Benchmarking radar plot of MEFET vs. other technologies.}\vspace{-1em}
% \label{radar}
% \end{center}
% \end{figure}

\begin{figure*}[h]
\begin{center}
\begin{tabular}{cc}
\includegraphics [width=0.49\linewidth, height = 4cm]{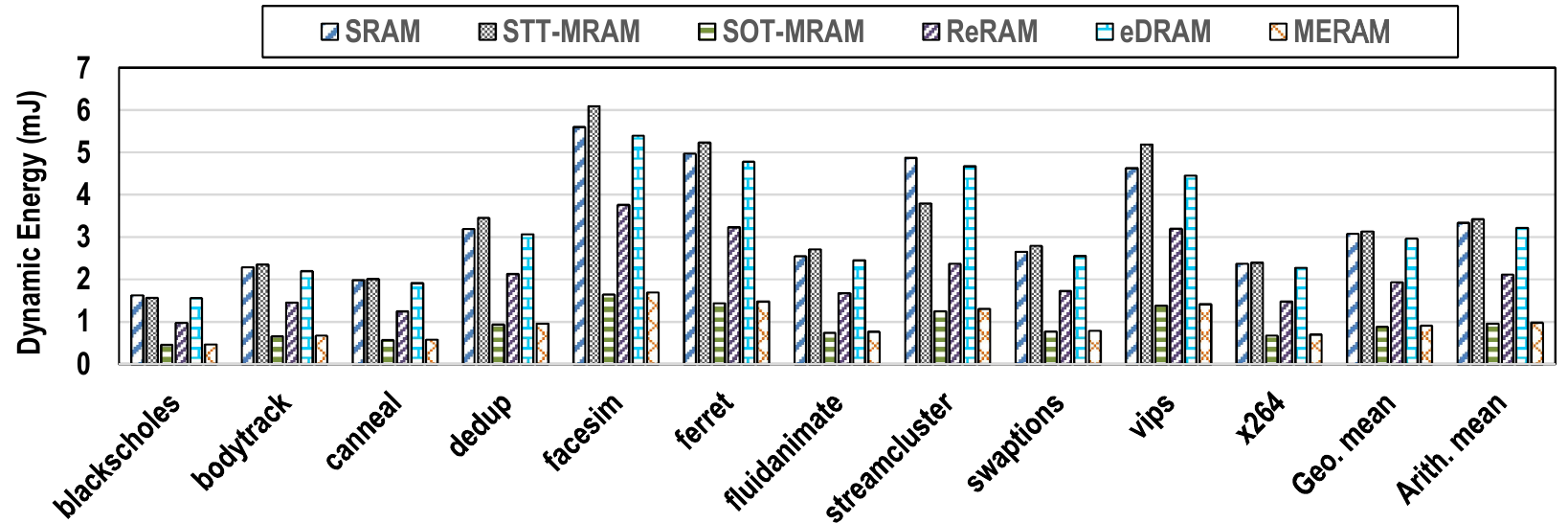} & \includegraphics [width=0.49\linewidth, height = 4cm]{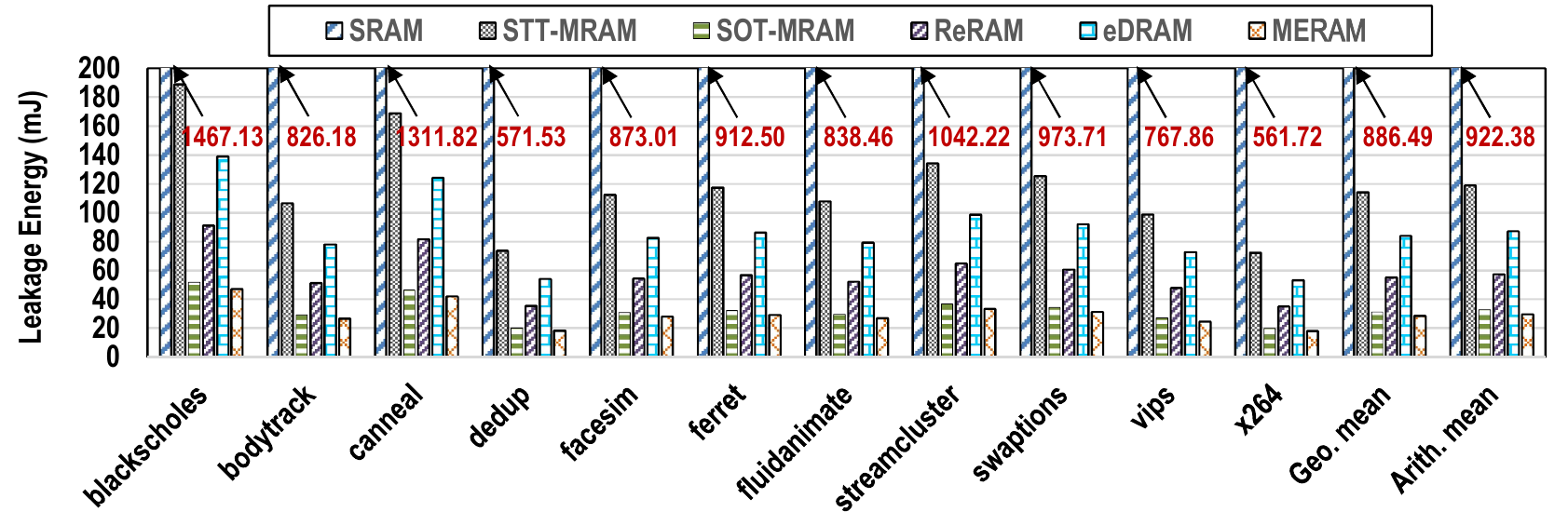} \vspace{-0.8em}\\
 \hspace{1.5 cm}   \small (a) &  \hspace{2.1 cm}  \small (b)\\\vspace{-0.4em}
 \end{tabular} \vspace{-1.4em}
\caption{L2 cache (a) dynamic energy and (b) leakage energy breakdown for SRAM, STT-MRAM, SOT-MRAM, ReRAM, eDRAM, and MERAM.}\vspace{-1.3em}
\label{Dynamic_energy}
\end{center}
\end{figure*}

\begin{figure*}[h]
\begin{center}
\begin{tabular}{cc}
\includegraphics [width=0.49\linewidth, height = 4cm]{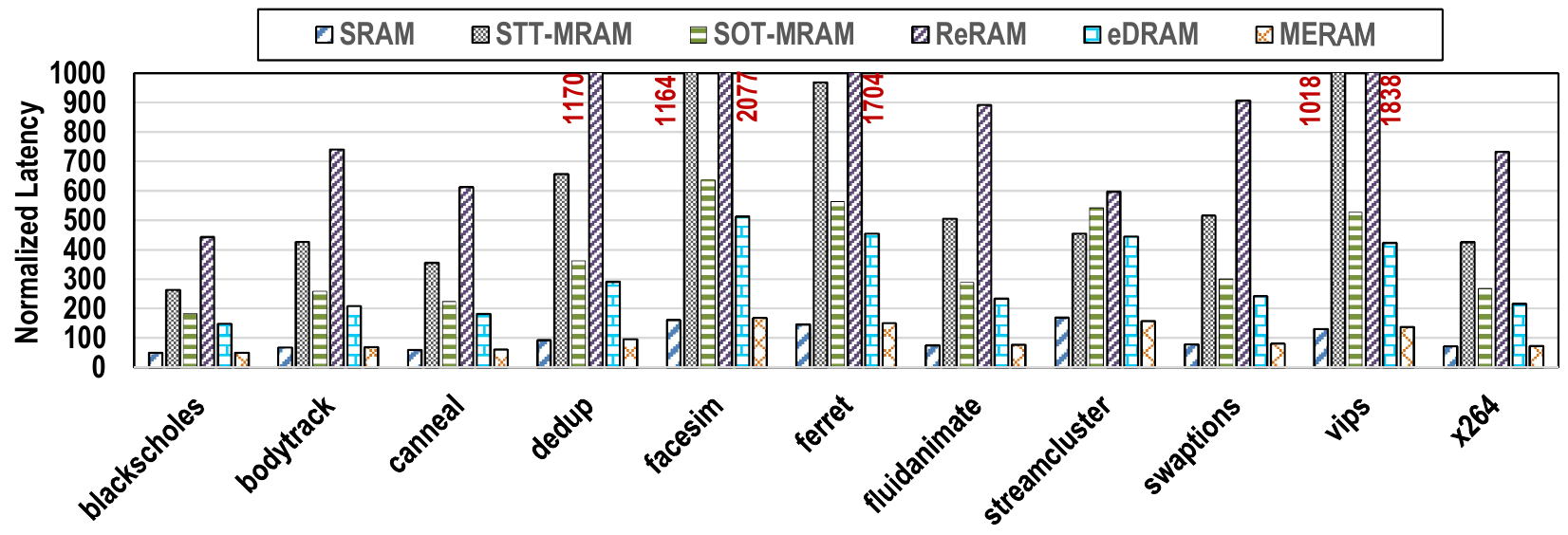} & \includegraphics [width=0.49\linewidth, height = 4cm]{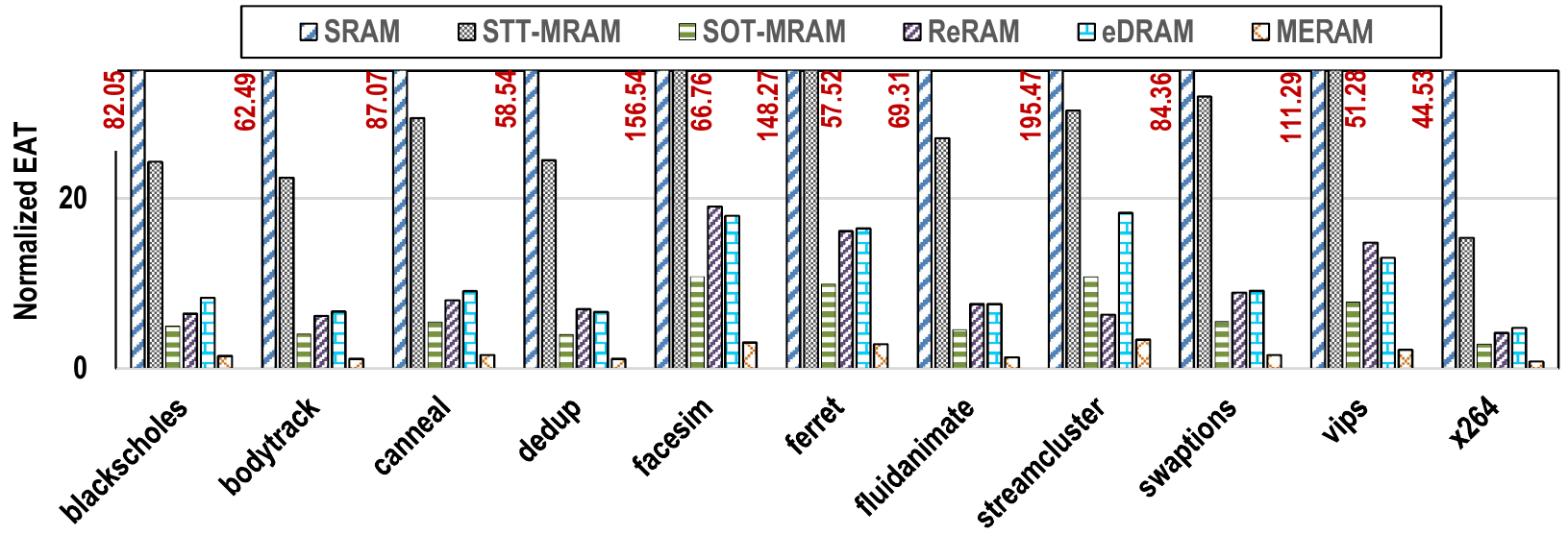} \vspace{-0.8em}\\
 \hspace{1.5 cm}   \small (a) &  \hspace{2.1 cm}  \small (b)\\\vspace{-0.4em}
 \end{tabular} \vspace{-1.4em}
\caption{(a) Latency comparison and (b) EAT comparison of various L2 candidates. The results are normalized w.r.t the average of latency (ns)/EAT for SRAM across all workloads and  among various L2 candidates.}\vspace{-1.3em}
\label{latency_arch}
\end{center}
\end{figure*}

\subsection{Architecture Level}

\subsubsection{Experiment Setup} 
The cycle-accurate simulator MARSSx86 \cite{patel2011marss} was used to evaluate the efficiency of our proposed circuit-to-architecture cross-layer framework. The cache controller modified to realize the functionality of L2 architecture with various memory technology candidates. We configured the simulator with the parameters listed in Table \ref{conf}. We selected eleven various benchmarks from PARSEC 2.1 suite for testing the performance of MERAM compared to the existing cache technologies.
The cache in the simulator warmed up with 5 million instructions.  The 500 million instructions starting at the Region Of Interest (ROI) of each workload was executed afterward. The collected reports were used to analyze the L2 candidates based on Energy Area Latency (EAT) product. The EAT metric can holistically identify the preferred L2 candidate by taking into account several essential metrics in the cache design.

\begin{table}[h]
\caption{System Configuration}\vspace{-1.5em}
\begin{center}
\scalebox{0.85}{
\begin{tabular}{|c|c|}
\hline
CPU         & 4 cores, 3.3 GHz, Fetch/Exec/ Commit width 4            \\ \hline
L1          & private, 32 KB, I/D separate, 8-way, 64 B, SRAM, WB \\ \hline
L2          & private, 4 MB, unifed, 8-way, 64 B, memory tech. candidate, WB      \\ \hline
Main Memory & 8 GB, 1 channel, 4 ranks/ channel, 8 bank/ rank         \\ \hline
\end{tabular}}
\end{center}
\label{conf}
\end{table}

\subsubsection{Energy Comparison}
Fig. \ref{Dynamic_energy}a shows the dynamic energy consumption comparison among L2 candidates. The dynamic energy consumption varies based on the workload characteristics (e.g. read/write intensity), and the power required for read/write operation in the given memory technology. In particular, the workloads with higher $\frac{write}{read}$ ratio impose considerably more dynamic energy in the candidates with high write energy consumption such as STT-MRAM. %Among the selected benchmarks, the {\fontfamily{cmtt}\fontsize{8}{7.2}\selectfont vips} and {\fontfamily{cmtt}\fontsize{8}{7.2}\selectfont facesim} have the highest $\frac{write}{read}$ ratio than other benchmarks.
%Hence, the PCM-based L2 candidate consumes the highest dynamic energy while running write intensive workloads. 
As an example, the dynamic energy consumption for heavily write intensive {\fontfamily{cmtt}\fontsize{9}{7.2}\selectfont facesim} and {\fontfamily{cmtt}\fontsize{9}{7.2}\selectfont ferret} are 6.09 mJ and 5.22 mJ, respectively in STT-MRAM based L2 candidate which are considerably higher than other candidates. On the other hand, the read intensive workloads such as {\fontfamily{cmtt}\fontsize{9}{7.2}\selectfont streamcluster} experience relatively higher number of read accesses. Thus, running the read intensive workloads incurs considerable high dynamic energy consumption in the candidates with relatively energy-costly read access. Among the L2 candidates, the SOT-RAM and MERAM consume the least dynamic power compared to other candidates due to leveraging innovative approaches to control the magnetic state of memory cell.

The execution time of each workload along with the leakage power unit for each L2 candidate were used to compute the leakage energy. As it is noticeable in Fig. \ref{Dynamic_energy}b, the SRAM incurs significantly higher leakage power, primarily due to its subthreshold leakage paths and the gate leakage current. Among the L2 candidates, the MERAM consumes the least leakage energy due to its intrinsically energy-efficient operations. Since the leakage energy is the major contributor to the overall energy consumption, the low leakage power consumption can significantly reduce the corresponding EAT for that individual L2 candidate.

% \begin{figure}[b]
% \begin{center}
% \begin{tabular}{c}
% \includegraphics [width=0.99\linewidth, height = 3.5cm]{./Figures/Leakage_arch_crop.pdf}\vspace{-0.4em}
%  \end{tabular} \vspace{-1em}
% \caption{L2 leakage energy breakdown for SRAM, STT-MRAM, SOT-MRAM, ReRAM, eDRAM, and MERAM.}\vspace{-1.3em}
% \label{Leakage_energy}
% \end{center}
% \end{figure}

\subsubsection{Latency Comparison}
We utilized the key parameters acquired from NVSim, CACTI, and our MERAM memory library, to compute the overall access latency for read/write operations based on the cache access pattern for each workload, as illustrated in Fig. \ref{latency_arch}a. To reduce the standard deviation of the simulation results w.r.t the commercialized design, we integrated the acquired profiles with the latency associated with the peripheral circuits. The SRAM offers a rapid read/write access compared to other L2 candidates because of its symmetrical structure that enables easily detectable minor voltage swings. The MERAM is the closest candidate to SRAM which offers significantly low read/write latency. This means that SRAM and MERAM should benefit from their low overall latency while estimating EAT for each L2 candidate.

% \begin{figure}[t]
% \begin{center}
% \includegraphics [width=0.99\linewidth, height = 3.5cm]{./Figures/Latency_crop.pdf}\vspace{-0.4em}
% \vspace{-1.7em}
% \caption{Latency comparison of various L2
% candidates. The results are normalized w.r.t the average of latency (ns) for SRAM across all workloads.}
% \vspace{-1.8em}
% \label{latency_arch}
% \end{center}
% \end{figure}

\subsubsection{EAT Product}
The MERAM offers an SRAM-competitive performance, superior energy consumption, and admissible area overhead compared to other candidates which makes it preferred L2 candidate. As illustrated in Fig. \ref{latency_arch}b, the MERAM delivers the least EAT among the L2 candidates. Compared to SOT-RAM which was considered as the superior alternative, MERAM reduces the EAT by 70.81\% on average. In the heavily write intensive workloads like {\fontfamily{cmtt}\fontsize{9}{7.2}\selectfont vips} and {\fontfamily{cmtt}\fontsize{9}{7.2}\selectfont facesim}, the EAT is reduced by 71.74\% and 71.44\%, respectively relative to delivered EAT in SOT-RAM. This trend is also seen in the read intensive {\fontfamily{cmtt}\fontsize{9}{7.2}\selectfont streamcluster} workload, whereby the EAT reduced by 68.63\% w.r.t delivered EAT in SOT-MRAM candidate.  To be specific, MERAM decreases the EAT by 80.26\%, 82.48\%, 94.57\%, and 98.12\% w.r.t ReRAM, eDRAM, STT-MRAM, and SRAM, respectively.

% \begin{figure}[h]
% \begin{center}
% \includegraphics [width=0.99\linewidth, height = 3.5cm]{./Figures/EAT_crop.pdf}\vspace{-0.4em}
% \vspace{-1.5em}
% \caption{EAT comparison among various L2
% candidates. The results are normalized w.r.t the average of EAT for SRAM across all workloads.}
% \vspace{-1.8em}
% \label{EAT}
% \end{center}
% \end{figure}

\section{Conclusions} 
In this work, we presented a non-volatile 2T-1MEFET memory bit-cell with separate read and write paths. We designed a device-to-architecture cross-layer evaluation framework to quantitatively analyze and compare the proposed design with other memory architectures. Our simulation results showed  that MERAM offers an SRAM-competitive performance, superior energy consumption, and admissible area overhead compared to other candidates which makes it the preferred L2 candidate. 
As an L2 cache alternative, MERAM reduces Energy Area Latency (EAT) product on average by $\sim$98\% and $\sim$70\%  relative to 6T SRAM and 2T SOT-MRAM platforms, respectively. We believe such circuit/architecture experiments can bring important motivation and guidance to device-level researchers in this domain to see the potential performance of this new emerging paradigm.
\vspace{-0.7em}

% \section{Acknowledgement}
% This project was partially supported by nCORE, a wholly owned subsidiary of the Semiconductor Research Corporation (SRC), through task ID-2968.001, 2760.001, 2760.002, and by the National Science Foundation (NSF), through Grant No. NSF-CCF 2005209, NSF-ECCS 1740136.

\bibliographystyle{IEEEtran}
{\tiny \bibliography{IEEEabrv,./Manuscript.bib}}

\end{document}